# Comparative Evaluation of the Proximate and Cytogenotoxicity of Ash and Rice Chips Used as Mango Fruit Artificial Ripening Agents in Birnin Kebbi, Nigeria

*Obadiah CD, Yahaya TO, Aliero AA, Abdulkareem M.

Department of Biological Sciences, Federal University Birnin Kebbi, P.M.B 1157, Birnin Kebbi, Kebbi State, Nigeria.

*Correspondence: calebdikko@gmail.com; Obadiah.caleb@fubk.edu.ng; 08038509764

**Abstract**

The high demand for mango (*Mangifera indica* L.) fruits has led sellers to employ ripening agents. However, concerns are growing regarding the potential toxicities of induced ripening, emphasizing the need for scientific investigation. Samples of artificially and naturally ripened mangoes were analyzed for proximate composition using standard protocols. Cytogenotoxicity was then assessed using the *Allium cepa* L. toxicity test. Twenty (20) *A. cepa* (onion) bulbs were used, with 5 ripened naturally, 5 with wood ash, 5 with herbaceous ash, and 5 with rice chips, all grown over tap water for five days. The root tips of the bulbs were assayed and examined for chromosomal aberrations. The results revealed a significant ($P<0.05$) increase in moisture, protein, and ash content of mangoes as ripening agents were introduced. Mangoes ripened with wood ash exhibited the highest moisture content (81%), while those ripened with rice chips had the highest protein (0.5%) and ash content (1.5%). Naturally ripened mangoes displayed the highest fat (0.0095%) and fiber (11.46%) contents. The *A. cepa* toxicity test indicated significant ($p<0.05$) differences in the root growth of mangoes ripened with various agents. Wood ash resulted in the highest root growth (2.62cm), while herbaceous ash had the least (2.18%). Chromosomal aberrations, including sticky, vagrant, and laggard abnormalities, were observed in all agents, with herbaceous ash exhibiting the highest and rice chips the least. The obtained results suggest that induced ripening of the fruits could induce toxicities, highlighting the necessity for public awareness regarding the potential dangers posed by these agents.

**Keywords:** Ash, Chromosomal aberrations, Fruit ripening, Mango, and Rice chips

## 1.0 Introduction

Fruit ripening is a complex process that influences the pigments, sugar, acid, flavor, aroma, texture, and color, making fruits appealing and suitable for consumption [4]. It is a highly regulated and programmed developmental phenomenon involving the coordination of multiple metabolic changes, as well as the activation and inactivation of various genes, resulting in biochemical and physiological changes within the tissue [18].

Fruits are classified into two classes based on the regulatory processes leading to ripening: climacteric and non climacteric fruits [15]. Mangoes, bananas, guavas, papayas, apples, and melons are examples of climacteric fruits, which are ethylene-dependent and can ripen after harvest, often with the help of exogenous ethylene. Non climacteric fruits ripen only if they remain attached to the parent plant. Examples include grapes, pineapples, strawberries, and citrus fruits [8, 15, 7].

Mango (*Mangifera indica* L.) is a climacteric fruit tree crop belonging to the genus *Mangifera* of the family Anacardiaceae, widespread across the tropics and subtropics [14]. For years, the fruit has been produced for its rich composition of phytonutrients and has been incorporated into the human diet as a source of carbohydrates, proteins, vitamins (particularly vitamins A and C), beta-carotene, minerals, and bioactive compounds such as phenolic acids, polyphenols, and carotenoids [6. 21]. The fruit is commonly harvested at the green, unripe stage, allowing for transportation and sale to distant markets [20]. As the demand for fresh mango fruit grows, the challenge of producing and distributing consumer-safe commodities in adequate and reasonable quantities becomes significant. Several artificial techniques and additives have been employed to induce rapid post-harvest ripening of the fruit [13, 18, 7].

During the artificial ripening process, alterations in the physio-chemical composition [5] and possibly genetic damage in the cells of mango fruit may occur. Unfortunately, a literature search shows a dearth of documented information on the safety of ripening agents in Birnin Kebbi, Nigeria. This information is necessary to raise public awareness if found toxic. Therefore, the current research aims to evaluate the proximate composition and cytogenotoxicity of wood ash, herbaceous ash, and rice chips used as artificial ripening agents for mango fruits in Birnin Kebbi.

## 2.0 Materials and Methods

### 2.1 Study area

The study was conducted at Kali Farm, situated in the Aliero Local Government Area of Kebbi State, Nigeria. The local government is located in the southeastern part of Kebbi State at latitude 12.27709 and longitude 4.44493. The town is well-known for its production of various vegetables and fruits, including groundnuts, pepper, tomatoes, onions, mangoes, cashews, and watermelons. Additionally, other food crops grown in this area include sorghum, beans, rice, and sugarcane [17]. In general, the land in the study area is extensively used for agricultural and settlement purposes. Kali Farm covers an area of 538,195.521 square feet, equivalent to five hectares of land. It is situated at latitude 12° 17' 24.14"N and longitude 4° 28' 1.57"E within the Aliero Local Government Area of Kebbi State, Nigeria [17].

The prevailing climate in the study area is categorized as tropical wet and dry. In a typical year, the average annual rainfall in and around this region amounts to approximately 800mm – 1000mm. The temperature remains consistently warm or hot throughout the year, with a slight cooling period occurring between November and February. Among these months, December is considered the coolest, with mean values ranging between 12°C and 21°C. On the other hand, April and May are the hottest months, with average mean temperatures reaching 35°C. Temporal variations in temperature from year to year are minimal. Actual evapotranspiration is estimated to be around 75% of the rainfall in the southern part of the region, increasing to approximately 80% in the northern part. Relative humidity experiences a 5% increase in the south and a corresponding decrease in the north [10].

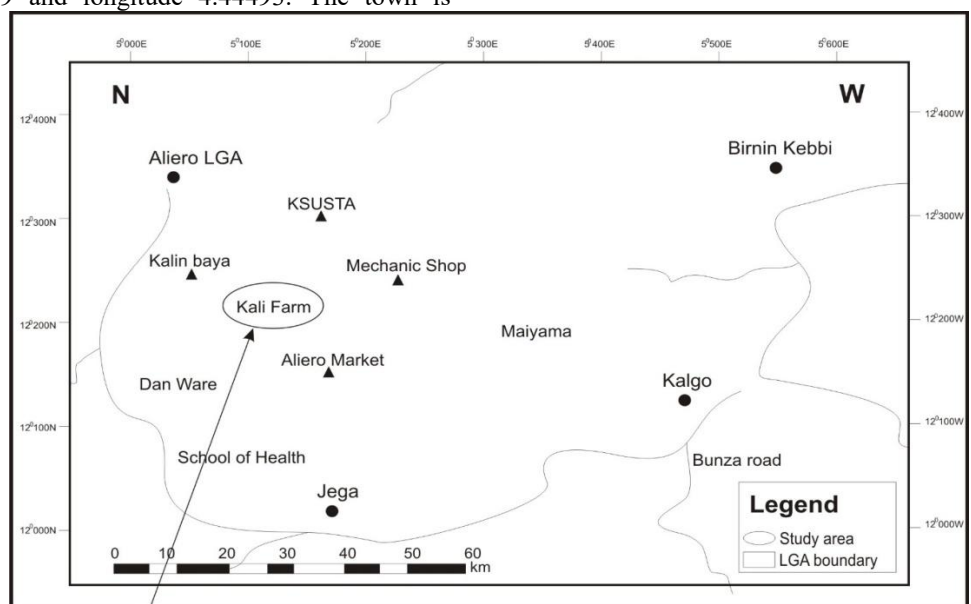

**Fig 2:** Map showing Kali Farm and its environs.
**Source**: GIS unit, FUBK, 2022

### 2.2 Sample collection

Two naturally ripen and six fully mature unripe mango fruits and were collected from Kali farm in Aliero local government. They were then transported to the laboratory of the Department of Biological Sciences at the Faculty of Science, Federal University, Birnin Kebbi, Kebbi State for artificial ripening using ash and rice chips; it should be noted that only undamaged and free from visible signs of infection mangoes were selected and plucked directly from the tree.

### 2.3 Experimental Design

The unripe fruits were randomly selected and grouped into three groups with each group having two mangoes replicates. In all the three groups, each mango was ripened separately in a polythene bag (two polythene bags for each group). The first group were ripened in wood ash, the second group were ripened in herbaceous ash and the third group were ripened in rice chips. On attaining the ripening stage indicated by change in flesh colour (from green to pale yellow) and indented upon pressing with the thumb [16] they were then processed for proximate analysis. The ripening agents were also kept for further cytogenotoxicity examination.

### 2.4 Methodology

The standard analytical method of the Association of Official Analytical Chemists, [2] was used to determine the moisture, ash, fibre, protein, fat and carbohydrate contents of the mango samples (dry matter). The cytogenotoxic effects of the ripening agents was evaluated using the methods proposed by [3].

#### 2.4.1 Determination of moisture content

A crucible with no contents was measured, and a gram of mature mangoes was measured into the crucible, and the weight of the crucible and sample was recorded; the sample was then dried in the oven overnight and subsequently removed and cooled in the desiccators. The sample was weighed after drying, the weight was recorded and was calculated by the following formula [2]:

Moisture content (%) $= \dfrac{w1 w2}{w1 - w0} \times 100$

Where: $w_0$ = weight of empty crucible; $w_1$ = weight of fresh sample + empty crucible; $w_2$ = weight of dried sample + empty crucible

#### 2.4.2 Determination of ash content

An empty crucible was weighed and one gram of the ripe mangoes sample was weighed into the crucible, the weight of the crucible and sample was taken. The sample was dried in the oven overnight, the sample was taken out after drying and was allowed to cool in the desiccator. The sample was weighed after drying, the weight was recorded and it was taken to the muffle furnace for ashing set at 550 $^o$C for 3 hours. It was kept at this temperature until white, light grey ash is obtained which appears to be free from carbonaceous particles. The crucible was placed in a desiccator and was allowed to cool. The weight of the sample was also taken after ashing and was calculated using the following formula [2].

Ash content (%) $= \dfrac{w1 - w2}{w1 - w0} \times 100\%$

Where: $w_0$ = weight of empty crucible; $w_1$ = weight of crucible + dry sample; $w_2$ = weigh of crucible + ash sample.

#### 2.4.3 Determination of fibre content

One gram of the ripe mangoes sample was placed into a beaker and 50ml of $H_2SO_4$ was added and boiled vigorously for exactly 30 minutes in the water bath. The sample was taken out, stirred and filtered with filter paper, the residue was transferred into the beaker and 50 ml NaOH was added into it. The sample was taken back to the water bath for digestion and was taken out after few minutes and re filtered. The residue was rinsed with HCL, Ethanol and Hot water respectively in other to dissolve the sample leaving only the fibre content of the sample. After the final filtration, the residue was washed into a weighed crucible and dried in the oven overnight. After drying, it was cooled in the desiccator and weighed rapidly. The sample was taken to the muffle furnace and ash at 550$^o$C for three hours. After heating, it was cooled in the desiccator before weighing and was calculated using the following formula [2].

Fibre (%) $= \dfrac{\text{loss in weight of ash on ignition}}{\text{original weight of sample}} \times 100$

#### 2.4.5 Protein quantification

The method is grounded on the fact that the digestion of a food sample with a strong acid such as $H_2SO_4$ results

Obadiah et al.,

in the release of nitrogen, which can be measured through titration, and the protein content can be estimated by multiplying the nitrogen concentration by a conversion factor of 6.25 [2].

$$\text{Crude protein (\%)} = \frac{(a-b) \times 0.01 MHCl \times 14c}{6.25 \times 100 \times d \times e}$$

Where: a = titre value for the digested sample; b = titre value for the blank; c = volume for which the digested was made up; d = volume of aliquot used in digestion; e = weight of dried sample

### 2.4.6 Determination of fat content

One gram of mango sample was weighed and transferred to an extraction thimble which was covered with a fat free wad of cotton wool. A clean and dry extraction flask was weighed to the nearest 1mg. The thimble was placed in an extractor and was extracted for 80 min with light petroleum. The residue was dried maintaining the flask for one hour in the drying oven at 98$^o$C. It was left to cool in a desiccator and weighed after cooling. The following formula was used to calculate the fat content [2]:

$$\text{Fat content (\%)} = \frac{\text{weight of extract}}{\text{weight of sample}} \times 100$$

### 2.4.7 Determination of carbohydrate content

The carbohydrate content was determined by the method of [2]:

%Total carbohydrate = (100- % (moisture + ash +protein + fibre + fat)

### 2.5 Cytogenotoxicity test

The cytogenotoxicity test was determine using *Allium cepa* (onion bulb)test following the method of [2]. Onion root tips was allowed to grow in tap water for five days and growth was observed. The onion root tips were later exposed to different dilutions of artificially ripening agents to observe the further growth. *A. cepa* was exposed for another five days to different dilutions of the ripening agents as follows:

  i. Wood ash: 5g/dl, 10g/dl, 15g/dl, 20g/dl, 25g/dl and the control
 ii. Herbaceous ash: 5g/dl, 10g/dl/dl, 15g/dl, 20g/dl, 25g/dl and the control
iii. Rice chips:5g/dl, 10g/dl, 15g/dl, 20g/dl, 25g/dl and control.

Each concentration was set-up in five replicates. The test solutions were stirred after every twenty-four hours. Two longest root tips were selected in each bulb and the root tips were measured after every twenty hours to observe any change in the growth. On reaching the fifth day, the root tips were cut and allowed to be soft in solution of 15ml acetic acid and 45ml ethanol. After the root tips dissolved, a slide was prepared using aceto-amine stain to stain the root and was observed under the microscope to view the cells and how the chromosomes are affected by those ripening agents [3].

### 2.6 Data Analysis

All data are presented as mean ± SD. Statistical significance at the 95% probability levels was set at P<0.05. Microsoft Excel (Version2016) was used for statistical analysis. Data was subjected to analysis of variance (ANOVA) and tested for significance difference among treatments.

### 3.0 Results

### 3.1 Proximate compositions of mangoes ripened naturally and artificially with ripening agents

Table 1 compares the proximate composition of mangoes ripened with rice chip, wood ash, and herbaceous ash with naturally ripened mangoes. There was a significant difference between the proximate parameters (moisture content, ash, protein, fat, fibre, and carbohydrate) of artificially ripened mangoes compared with naturally ripened. The mangoes ripened with wood ash was significantly (P<0.05) higher in moisture content (81%), while the highest protein (0.5%) and ash content (1.5%) were observed in mangoes ripened with rice chip. The ash content was found to be higher in rice chips (1.5%) than in both the naturally ripened mango (1.01%) and mangoes ripened with wood ash (0.35%) and herbaceous ash (0.8%). Fibre content was higher in naturally ripped mangoes (11.46%) than others. There was no significant difference (P<0.05) between the total carbohydrate content of naturally ripened mangoes (14.99%) and the mangoes artificially ripened with rice chips (15.4%) and herbaceous ash (14.95%). However, mangoes ripened with wood ash had the lowest carbohydrate content (9.92%). Additionally, naturally ripened mangoes significantly (P<0.05) had the highest (0.0095%) fat content, but there was no statistical (P<0.05) difference among the fat content of mangoes ripened with rice chip (0.005%), wood ash (0.005%), and herbaceous ash (0.0005%).

**Table 1: Proximate composition of mangoes ripened with rice chip, wood ash, herbaceous ash, and naturally ripened**

| Treatment | Moisture content | Ash | Fibre | Protein | Fat | Carbohydrate |
|---|---|---|---|---|---|---|
| Rice chips | $77.5 \pm 0.5^{ab}$ | $1.5 \pm 0.5^b$ | $5.1 \pm 0.1^a$ | $0.5 \pm 0.01^c$ | $0.005 \pm 0^{ab}$ | $15.4 \pm 0.89^b$ |
| Wood ash | $81 \pm 1^b$ | $0.35 \pm 0.05^a$ | $8.2 \pm 0.1^b$ | $0.53 \pm 0.05^c$ | $0.005 \pm 0^{ab}$ | $9.92 \pm 1.1^a$ |
| Herbaceous ash | $73 \pm 2^a$ | $0.8 \pm 0.1^{ab}$ | $8.4 \pm 0.2^b$ | $0.35 \pm 0^b$ | $0.0005 \pm 0^a$ | $14.95 \pm 0.61^b$ |
| Natural | $73 \pm 2^a$ | $1.01 \pm 0.21^{ab}$ | $11.46 \pm 1.04^c$ | $0.04 \pm 0.02^a$ | $0.0095 \pm 0^b$ | $14.99 \pm 0.26^b$ |

Values were presented as means ± S. E. M; values with superscripts a, b, or c (3 or more levels) are significantly (p<0.05) different (ANOVA)

### 3.2 Cytogenetic effects in mangoes ripened naturally and artificially with ripening agents

Table 2 shows the cytogenetic effects of wood ash, herbaceous ash, and rice chips on the growth of treated *A. cepa* bulbs. There were significant (p<0.05) differences in the root growth of the treated *A. cepa* after day one. Wood ash produced the highest root growth (2.62cm), while herbaceous ash produced the least (2.18cm). There were no significant differences in the root growth of the bulbs at day two, day three, and day four.

On the other hand, there was significant (p<0.05) decrease in the root growth of the bulbs as the weight of ash increases compared to the control across at all days. Additionally, there were significant differences in the root growth of bulbs exposed to different concentrations of ashes (5g, 10g, 15g, 20g and 25g) (Table 3).

**Table 2: Root growth of *Allium cepa* grown in different concentrations of ripening agents**

| Treatment | Initial root length (cm) | After day 1 (cm) | After day 2 (cm) | After day 3 (cm) | After day 4 (cm) |
|---|---|---|---|---|---|
| Wood ash | $2.27 \pm 0.13$ | $2.62 \pm 0.14^b$ | $2.99 \pm 0.36$ | $3.35 \pm 0.49$ | $3.56 \pm 0.6$ |
| Herbaceous ash | $2.03 \pm 0.15$ | $2.18 \pm 0.12^a$ | $2.32 \pm 0.15$ | $2.63 \pm 0.31$ | $2.74 \pm 0.38$ |
| Rice chip | $2.04 \pm 0.12$ | $2.31 \pm 0.12^{ab}$ | $2.37 \pm 0.11$ | $2.37 \pm 0.11$ | $2.37 \pm 0.11$ |

Group means and standard error are presented as; Means ± s. e. m: a, b, c, - means with different superscripts within factor (3 or more levels) are significantly (P>0.05) different

**Table 3: Root growth of *Allium cepa* grown in different concentrations of ripening agents**

| Weight of ash | Initial root length (cm) | After day 1 (cm) | After day 2 (cm) | After day 3 (cm) | After day 4 (cm) |
|---|---|---|---|---|---|
| 0g | $1.98 \pm 0.1$ | $2.85 \pm 0.24^b$ | $3.82 \pm 0.56^b$ | $4.87 \pm 0.71^b$ | $5.45 \pm 0.9^b$ |
| 5g | $1.85 \pm 0.22$ | $2.28 \pm 0.22^a$ | $2.3 \pm 0.21^a$ | $2.3 \pm 0.21^a$ | $2.33 \pm 0.22^a$ |
| 10g | $2.2 \pm 0.1$ | $2.32 \pm 0.12^{ab}$ | $2.38 \pm 0.12^a$ | $2.63 \pm 0.19^a$ | $2.67 \pm 0.21^a$ |
| 15g | $1.87 \pm 0.32$ | $1.98 \pm 0.19^a$ | $2.02 \pm 0.18^a$ | $2.02 \pm 0.18^a$ | $2.02 \pm 0.18^a$ |
| 20g | $2.33 \pm 0.06$ | $2.38 \pm 0.07^{ab}$ | $2.42 \pm 0.09^a$ | $2.45 \pm 0.11^a$ | $2.45 \pm 0.11^a$ |
| 25g | $2.43 \pm 0.15$ | $2.38 \pm 0.13^{ab}$ | $2.42 \pm 0.12^a$ | $2.42 \pm 0.12^a$ | $2.42 \pm 0.12^a$ |

Group means and standard error are presented as; Means ± s. e. m: a, b, c, - means with different superscripts within factor (3 or more levels) are significantly (p<0.05) different.

#### 3.2.1 Chromosomal aberations observed in the root cells of *Allium cepa* treated with ripening agents

The chromosomal abnormalities observed in the root cells of *A. cepa* treated with rice chips, wood ash, and herbaceous ash are presented in Plates 1-3. Rice chips induced bridge at telophase, stickiness at telophase, and vagrant chromosomes at metaphase (Plate 1); herbaceous ash induced stickiness at telophase and vagrant at metaphase (Plate 2); and wood ash caused stickiness at telophase, laggard chromosomes, and stickiness at telophase (Plate 3). The most frequent aberrations were bridges and sticky chromosomes, and herbaceous ash induced the highest number, followed by wood ash.

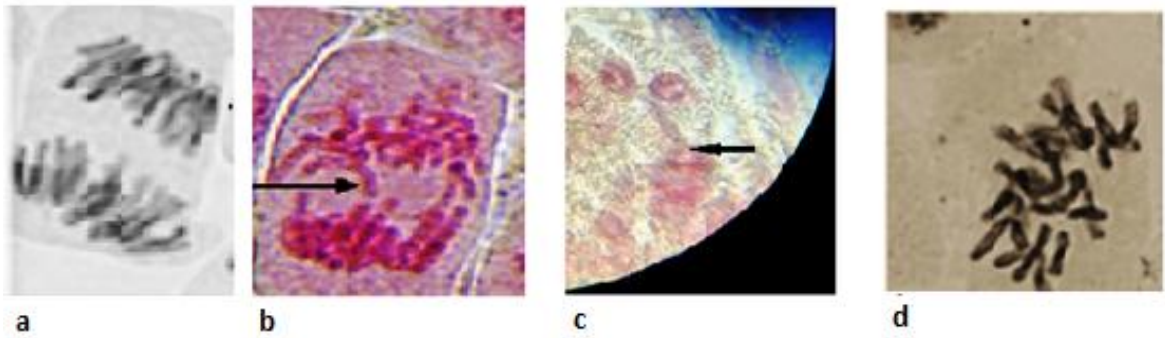

Plate 1: Chromosomal aberrations induced by rice chips: a = controlanaphase, b = chromosome bridge at telophase, c = stickiness at telophase, and d = vagrant at metaphase

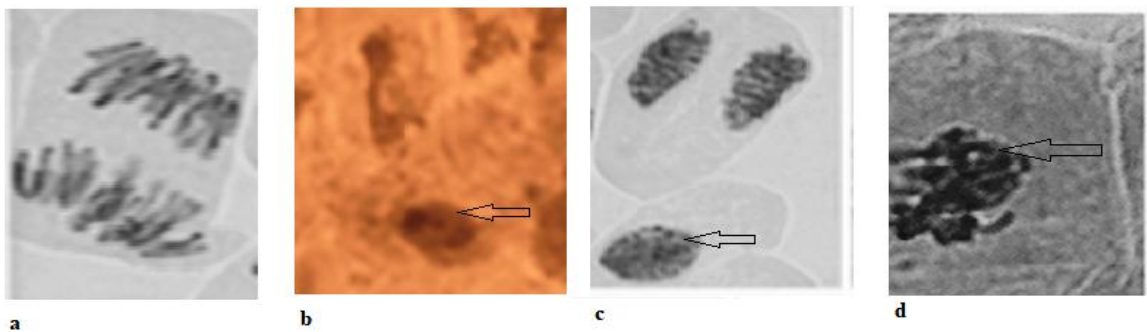

Plate 2: Chromosomal aberrations induced by herbaceous ash: a = control anaphase, b = stickiness at telophase, c = stickiness at telophase, and d = vagrant at metaphase

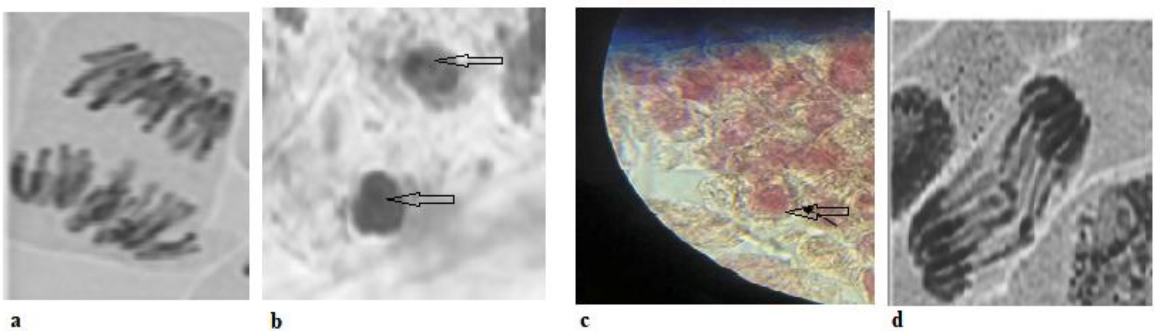

Plate 3: Chromosomal aberrations induced by wood ash: a = control anaphase, b = stickiness at telophase, c = laggard chromosomes, and d = stickiness at telophase

## 4.0 Discussion

The current study has unveiled the impact of ash and rice chips as artificial ripening agents on the proximate composition of mango fruits, comparing them with naturally ripened ones. This suggests that various artificial ripening agents bring about proximate alterations in mango fruits. Furthermore, the findings of the study illustrate that artificial ripening agents induce cytogenetic alterations in mango fruits, as evidenced by the *A. cepa* toxicity test.

### 4.1 Proximate compositions of mangoes ripened naturally and artificially with ripening agents

The moisture, ash, fiber, protein fat and carbohydrate content of mango increased as ripening agents were introduced (Table 1). High moisture content of mango ripened in wood ash may not be preferred by sellers because of the high moisture content (81%) which may lead to faster rate of microbial degradation and shorter shelf life resulting in quicker deterioration of the fruit[12,14]. This result is consistent with the finding of Ubwa et al. [19] for three mangoes varieties from Benue state, Nigeria.

The ash content of mango fruits is an index of their mineral content [19]. In the present study, ash content was found to be relatively higher in rice chips (1.5%) than both the naturally ripened mango (1.01%) and mangoes ripened in wood ash (0.35%) and herbaceous ash (0.8%). These findings indicate that the ripening agents affects the ash content of the fruit, which is important for nutrient content and further suggest that rich chips can improve the mineral content of the fruit.

There were some gradual decreases in the fibre content of the naturally ripped mangoes (11.46%) with the addition of ripening agents; herbaceous ash (8.4%), wood ash (8.2%) and rich chips (5.1%) respectively. The fiber content of the mangoes in this study were relatively high and higher than the NAFDAC minimum requirement of 3 g/100g for a source of fiber [11].

The protein content of mangoes ripened with rice chips observed in this study is relatively between the range of 1.65 and 1.3% reported for the three varieties of naturally ripened mango by Ubwa et al, [19].

Naturally ripened mangoes had the highest fat content (0.0095%), which was not statistically (p<0.05) different from the fat content of mangoes ripened with rice chip (0.005%) and wood ash (0.005%), while herbaceous ash had the least fat content (0.0005%). These results are much lower compared with the findings of Ubwa et al. [19] who recorded a fat content between the ranges of 0.06 – 0.16 %. The carbohydrate content recorded in this study is higher than those reported elsewhere. For example, Orisa et al. [13], reported that the carbohydrate content of the mango ranged from 2.27-5.24% with unripe mango recording the highest, while calcium carbide mango the lowest. Similarly, 19.78 – 19.60 % of carbohydrate content of naturally ripened mangoes was reported by [19]. The decrease in carbohydrate content during ripening is due to the hydrolysis of carbohydrates into sugars during ripening as suggested by the two studies. However, the forgoing findings on the higher carbohydrate content of mangoes ripened with rice chip, herbaceous ash and naturally ripped mangoes could suggest they are better categorized as energy sources than wood ash and calcium carbide.

**4.2 Cytogenotoxicity of artificial ripening agents**

All the ripening agents induced a reduced growth rate, along with chromosomal aberrations, including sticky chromosomes, laggard chromosomes, vagrant chromosomes, and chromosomal bridges. These results align with those of [23], who reported both up regulation and down regulation of genes in artificially ripened banana fruits. Alege and Anthony [1] also documented chromosomal abnormalities such as binucleate cells, vacuolated cells, sticky chromosomes, c-mitosis cells, and anaphase bridges in *Allium cepa* treated with calcium carbide, a fruit ripening agent. The formation of sticky chromosomes may be attributed to the intermingling of chromatin fibers and subsequent connection of sub-chromatids between chromosomes, while the creation of anaphase bridges could occur due to chromosome breakage and subsequent re-joining [1].

**5.0 Conclusions**

This study evaluates the proximate composition as well as the cytotoxic effects of wood ash, herbaceous ash, and rice chips used as artificial ripening agents for mango. The ripening agents significantly (P<0.05) increased the moisture, protein, and ash content of the mangoes. Mangoes ripened with wood ash had the least carbohydrate content. All the agents resulted in a reduced rate of growth, along with chromosomal abnormalities, including sticky chromosomes, laggard chromosomes, vagrant chromosomes, and chromosomal bridges. The herbaceous ash is the most toxic of all the ripening agents while rice chips is the least toxic.

**6.0 Recommendations**

The outcomes of this study imply that inducing ripening of the fruits could stimulate toxicities, emphasizing the requirement for public consciousness concerning the potential hazards presented by these agents.

## 7.0 Acknowledgments

The authors acknowledged the management of Kali farm, Aliero for supplying the mangoes fruit utilized in the research.

## 8.0 Conflict of Interest

The authors declared no conflict of interest.

## 9.0 References

[1] Alege GO, Anthony W. Assessment of Cytotoxicity Induced by Calcium Carbide (Fruit Ripening Agent) on *Allium Cepa* L. (Onion) Root Meristematic Cell:International Journal of Scientific Research in Biological Sciences. 2020; 7 (2): 170-175.

[2] AOAC. Association of Official Analytical Chemists. Official Methods of Analysis, (Vol.II, 17th edition) of AOAC international, Washington DC, USA. 2000;12.

[3] Bonciu E, Firbas P, Fontanetti CS, De Souza CP, Wusheng J, Liu D, Karaismailoğlu MC, Menicucci F, Pesnya DS, Romanovsky AV, Popescu A. An evaluation for the standardization of the *Allium cepa* test as cytotoxicity and genotoxicity assay:Caryologia. 2018;71(3):191-209. DOI : 10.1080/00087114.2018.1503496

[4] Ernesto DB, Omwamba M, Faraj AK,Mahungu, SM. Physico-chemical characterization of keitt mango and cavendish banana fruits produced in Mozambique: Food and Nutrition Sciences, 2018; 9(5): 556-571.

[5] Joshi H, Kuna A, Lakshmi MN, Shreedhar M, Kumar AK. Effect of stage of maturity, ripening and storage on antioxidant content and activity of *Mangiferaindica* L. var. Manjira. International Journal of Food Science and Nutrition.2017; 2(3): 1-9

[6] Kumar M, Saurabh V, Tomar, M, Hasan M, Changan S, Sasi M. *et al*. Mango (*Mangiferaindica* L.) Leaves: Nutritional Composition, Phytochemical Profile, and Health-Promoting Bioactivities:Antioxidants, 2021;10(2): 299. https://doi.org/10.3390/antiox10020299

[7] Lawaly MM. (2022). Effects of Calcium Carbide Used as a Fruit-Ripening Agent on Fruit Toxicity. European Journal of Nutrition and Food Safety. 2022; 64-76.

[8] Maduwanthi SD, Marapana RA. Induced ripening agents and their effect on fruit quality of banana: International journal of food science. 2019; 2;2019.

[9] Microsoft Corporation, USA Microsoft Excel 2019 for Windows. 2019. https://office.microsoft.com/excel

[10] Muhammad M, Singh D, Imonikhe AM, Keta M, Obaroh, I. Phytochemical analysis and ethnoecological survey of some species of African mistletoes collected from Birnin Kebbi, Kebbi state:Journal of Innovative Research in Life Sciences. 2022 4(1): 65–73.

[11] NAFDAC (National Agency for Food & Drug Administration & Control). 2013. Minimum Requirement for Analysis of Finished Product. Summary of Current Food standards as of 04 April, 2013.

[12] Odion EE, Nwaokobia K, Ogboru RO, Olorode EM, Gbolagade OT. Evaluation of some Physico-chemical Properties of Mango (*MangiferaIndica* L.) Pulp in South-south, Nigeria. Ilorin Journal of Science. 2020 Dec 1;7(2):236-48.

[13] Orisa CA, Usoroh CI, Ujong AE. Accelerated Ripening Agents and Their Effect on the Quality of Avocado (*Persiaamericana* M.) and Mango (*Mangiferaindica* L.) Fruits. Asian Journal of Advances in Agricultural Research. 2020;29–40. https://journalajaar.com/index.php/AJAAR/article/view/277

[14] Pamela EA, Usifo GA, Kikelomo OE, Yemisi OO, Omolara IA, Ayodeji OA. Diversity of Mango (*Mangifera Indica* L.) Cultivars Based on Physicochemical, Nutritional, Antioxidant, and Phytochemical Traits in South West Nigeria, International Journal of Fruit Science. 2022; 20(2): 352-376, DOI: 10.1080/15538362.2020.1735601

[15] Premarathne RMDE, Marapana RAUJ, Perera PRD. Determination of Physiochemical Characteristics and Antioxidant Properties of Selected

Climacteric and Non–climacteric Fruits. Journal of Pharmacognosy and Phytochemistry. 2021; 10(3): 23-28.

[16] Raghavendra A, Guru D, Rao, MK, Sumithra R. Hierarchical approach for ripeness grading of mangoes. Artificial Intelligence in Agriculture. 2019; 4:243-252. https://doi.org/10.1016/j.aiia.2020.10.003

[17] Singh D, Jibril NK, Muhammad A, Malik AI, Osesua BA. Phytoremediation Potential of *Jatropha curcas* and *Cassia occidentalis* on Selected Heavy Metals in Tie Soil: JAP. 2022;5(1):33-49. Available from: https://www.carijournals.org/journals/index.php/JAP/article/view/864

[18] Singh P, Tarkha A, Kumar P, Singh J. Impact of Chemicals on the Ripening Physiology of Fruits. IntJCurrMicrobiolAppSci. 2020;9(12):219–29. https://www.ijcmas.com/abstractview.php?ID=20424&vol=9-12-2020&SNo=29

[19] Ubwa ST, Ishu MO, Offem JO, Tyohemba RL, Igbum GO. Proximate composition and some physical attributes of three mango (*Mangiferaindica* L.) fruit varieties:International Journal of Agronomy and Agricultural Research (IJAAR) 2014; 4(2): 21-29. ISSN: 2223-7054 (Print) 2225-3610 (Online)\

[20] Vanoli M, Rizzolo, A, Grassi, M, Spinelli L, Torricelli A. Modelling mango ripening during shelf life based on pulp color non-destructively measured by time-resolved reflectance spectroscopy. Scientia Horticulturae, 2023; 310: 111714.

[21] Yahaya T, Oladele E, Sifau M, Audu G, Bala J, Shamsudeen A. Characterization and Cytogenotoxicity of Birnin Kebbi Central Abattoir Wastewater. *Uniport J Eng Sci Res,* 2020; 5 (special issue): 63-70.

[22] Yahia EM, de Jesús Ornelas-Paz J, Brecht JK, García-Solís P, Celis ME. The contribution of mango fruit (*Mangiferaindica* L.) to human nutrition and health. Arabian Journal of Chemistry. 2023; 29:104860.

[23] Yan H, Wu F, Jiang G, Xiao L, Li Z, Duan X, Jiang Y. Genome-wide identification, characterization and expression analysis of NF-Y gene family in relation to fruit ripening in banana. Postharvest Biology and Technology: 2019;151: 98-110. https://doi.org/10.1016/j.postharvbio.2019.02.002.